\documentclass[pre,aps,showpacs]{revtex4}
\usepackage{epsf}
\usepackage{amsmath}

\def\nn{\nonumber}
\def\beq{\begin{equation}}
\def\eeq{\end{equation}}
\def\bea{\begin{eqnarray}}
\def\eea{\end{eqnarray}}
\def\bi{\begin{itemize}}
\def\ei{\end{itemize}}
\def\be{\begin{enumerate}}
\def\ee{\end{enumerate}}

\begin{document}

\title{Selfduality for coupled Potts models on the triangular lattice}
\author{Jean-Fran\c{c}ois Richard${}^{*,\dagger}$}
\author{Jesper Lykke Jacobsen${}^*$}
\author{Marco Picco${}^\dagger$}
\affiliation{${}^*$LPTMS, Universit\'e Paris-Sud, B\^atiment 100, 91405 
Orsay, France}
\affiliation{${}^\dagger$LPTHE, Universit\'e Paris VI, Tour 16,
%         Bo\^{\i}te 126, Tour 16, 1$^{\it er}$ {\'e}tage,
         4 place Jussieu, 75252 Paris cedex 05, France}
\date{\today}

\begin{abstract}
We present selfdual manifolds for coupled Potts models on the triangular
lattice. We exploit two different techniques: duality followed by decimation,
and mapping to a related loop model. The latter technique is found to be
superior, and it allows to include three-spin couplings. Starting from
three coupled models, such couplings are necessary for generating selfdual
solutions. A numerical study of the case of two coupled models leads to the
identification of novel critical points.
\end{abstract}

% 05.50.+q Lattice theory and statistics (Ising, Potts, etc.)
% 05.20.-y Classical statistical mechanics

\maketitle

\section{Introduction}
\label{sec1}

The two-dimensional Potts model is a well-studied model of statistical
mechanics \cite{Wu82} and continues to attract the interest of many workers.
Its definition is simple. Given a lattice with vertices $\{i\}$ and edges
$\langle ij \rangle$, the Hamiltonian reads
\beq
 \beta H = -K \sum_{\langle ij \rangle} \delta(S_i,S_j),
\eeq
where $\delta$ is the Kronecker delta.
The spins $S_i=1,2,\ldots,q$ initially take $q$ discrete values.
However, by making a random cluster expansion \cite{Kasteleyn} it is easily
seen that the partition function can be written
\beq
 Z = \sum_{\cal C} b^{e({\cal C})} q^{n({\cal C})},
 \label{ZPotts}
\eeq
where $b={\rm e}^K-1$. Here, the sum is over the $2^{|\langle ij \rangle|}$
possible colourings ${\cal C}$ of the edges (each edge being either coloured
or uncoloured), $e({\cal C})$ is the number of coloured edges, and $n({\cal
C})$ is the number of connected components (clusters) formed by the coloured
edges. Taking Eq.~(\ref{ZPotts}) as the {\em definition} of the Potts model,
it is clear that $q$ can now be considered as a real variable, independently
of the original spin Hamiltonian. Also, we shall adopt the point of view that
Eq.~(\ref{ZPotts}) makes sense for any real $b$, although $b<-1$ would
correspond to an unphysical (complex) value of the spin coupling $K$.

Exact evaluations of Eq.~(\ref{ZPotts}), in the sense of the Bethe Ansatz,
exist for several lattices and for specific curves in $(q,b)$ space along
which the model happens to be integrable \cite{BaxterBook}. This is true,
in particular, for the square lattice with \cite{BaxterBook,Baxter82}
\bea
 b &=& \pm \sqrt{q}, \label{sqFM} \\
 b &=& -2 \pm \sqrt{4-q}, \label{sqAF}
\eea
and for the triangular lattice with \cite{Baxter78}
\beq
 b^3 + 3 b^2 = q. \label{trisol}
\eeq
%
% Also true for triangular lattice with b=-1, but under duality this
% maps to something with nonzero 3-spin coupling. Also, I'm not so sure
% how to interpret this line in CFT.
%
These curves have several features in common. First, they correspond to
critical points (with correlation functions decaying as power laws)
for $0 \le q \le 4$ \cite{Baxter73}, whose nature can be classified using
conformal field theory (CFT) \cite{Saleur91}. Second, the values of the
coupling constants are often so that the partition function is selfdual
(see below); this is the case for the curves (\ref{sqFM}) and (\ref{trisol})
above, whereas the two curves in (\ref{sqAF}) are mutually dual.

The part of the curves having $b>0$ corresponds to the ferromagnetic phase
transition, whose critical behaviour is lattice independent (universal). More
interestingly, the antiferromagnetic ($-1 \le b < 0$) and unphysical regimes
($b < -1$) contain non-generic critical points whose relation to CFT has,
at least in some cases, not been fully elucidated. This is so in particular
for $b=-1$, where the Potts model reduces to a colouring problem, and
Eq.~(\ref{ZPotts}) becomes the chromatic polynomial.

Much less is known about several Potts models, coupled through their
energy density $\delta(S_i,S_j)$. Results coming from integrability seem to be
limited to the case of $N=2$ coupled models \cite{Vaysburd}, which on the
square lattice only leads to new critical points in the well-studied
Ashkin-Teller case \cite{BaxterBook} (i.e., with $q=2$). Apart from that,
CFT-related results are essentially confined to perturbative expansions in
$\epsilon \sim q-2$ around the ferromagnetic critical point
\cite{Cardy87,Ludwig87}. These results, corroborated by numerical evidence
\cite{Dotsenko99,Jacobsen00}, indicate the existence of novel critical
points for $N \ge 3$, with possible implications for the random-bond
Potts model through the formal analytical continuation (replica limit)
$N\to 0$.

In the present publication we investigate the possibility of novel critical
behaviour in $N$ coupled Potts models on the triangular lattice. To identify
candidate critical points we first search for selfdual theories. In comparison
with a similar study on the square lattice \cite{Dotsenko99,Jacobsen00}
several distinctive features emerge due to the non-selfdualness of the
lattice. This leads us to use two different techniques. In the first, a
standard duality transformation is followed by decimation (star-triangle
transformation). This turns out to be quite cumbersome, already for $N=2$. We
therefore turn to a second technique, which utilises a mapping to a system of
coupled loop models. This leads to simpler relations, and as a bonus allows
to include three-spin couplings around one half of the lattice faces.
Starting from $N=3$ coupled models, such additional couplings are actually
necessary for generating non-trivial selfdual solutions.

For $N=2$ we numerically investigate the non-trivial selfdual
manifold.  Following the motivation given above, the main interest
here is to establish whether a given selfdual point corresponds to a
renormalisation group fixed point (and possibly even to a critical fixed
point). We shall see that these expectations are indeed born out: the
numerics is compatible with critical points whenever $0 \le q \le
4$. Measuring the central charge, we identify the corresponding
universality classes. These can in some cases be understood from those
of a single model, but we also identify points possessing novel
critical behaviour.

The paper is laid out as follows. In Section~\ref{sec2} we present the
technique of duality followed by decimation for two coupled models with pure
two-spin interactions. In particular, we find a non-trivial selfdual solution.
The mapping to a loop model, given in Section~\ref{sec3}, allows to rederive
this solution in a much simpler way, and to generalise to the case where
three-spin interactions are included. In Section~\ref{sec4} we use this
technique to treat the case of three coupled models with both two and
three-spin interactions. A numerical study of the non-trivial selfdual
solution found in Section~\ref{sec2} is the object of Section~\ref{sec5}.
Finally, Section~\ref{sec6} is devoted to our conclusions.

\section{Models with two-spin interactions}
\label{sec2}

To illustrate the first technique (duality and decimation), we consider the
case of $N=2$ coupled models with two-spin interactions. In order to simplify
the notation, we introduce the symbol $\delta^{\mu}_{ij} = \delta \left(
S^{\mu}_i,S^{\mu}_j \right)$, where the superscript refers to the spins of the
$\mu$'th model ($\mu=1,2,\ldots,N$). We are interested in the coupled model
defined by the Hamiltonian
\beq
 \beta H_2 = - \sum_{\langle ij \rangle} \left \lbrace
 K_1 \delta^1_{ij} + K_2 \delta^2_{ij} + K_{12} \delta^1_{ij} \delta^2_{ij}
 \right \rbrace.
 \label{H2}
\eeq
The spins $S^\mu_i$ take $q_\mu$ different values.

\subsection{Duality followed by decimation}

As shown in Ref.~\cite{Jacobsen00,Dotsenko02}, Eq.~(\ref{H2}) admits a
(generalised) random cluster expansion resulting in
\beq
 Z = \sum_{{\cal C}_1,{\cal C}_2}
 b_1^{e({\cal C}_1 \cap \overline{{\cal C}_2})}
 b_2^{e(\overline{{\cal C}_1} \cap {\cal C}_2)}
 b_{12}^{e({\cal C}_1 \cap {\cal C}_2)}
 q_1^{n({\cal C}_1)} q_2^{n({\cal C}_2)},
 \label{Z2}
\eeq
where ${\cal C}_\mu$ are independent colourings of the $\mu$'th model,
and we have defined the complementary colouring
$\overline{{\cal C}_\mu} \equiv \langle ij \rangle - {\cal C}_\mu$.
The new parameters $b$ are related to the coupling constants $K$ through
\beq
 b_\mu = {\rm e}^{K_\mu}-1, \qquad
 b_{12} = {\rm e}^{K_1+K_2+K_{12}}-{\rm e}^{K_1}-{\rm e}^{K_2}+1 \label{defb}
\eeq
As explained in the Introduction, we shall take the point of view that
the model is {\em defined} by Eq.~(\ref{Z2}) for any real values of
$b$ and $q_\mu$.

Up to an irrelevant constant, the partition function of the dual model
is again given by (\ref{Z2}), but now with respect to the dual (hexagonal)
lattice, and with dual values $\tilde{b}$ of the parameters
\cite{Jacobsen00,Dotsenko02}:
\beq
 \widetilde{b_1}=\frac{b_2\,q_1}{b_{12}}, \qquad
 \widetilde{b_2}=\frac{b_1\,q_2}{b_{12}}, \qquad
 \widetilde{b_{12}}=\frac{q_1\,q_2}{b_{12}}.
 \label{dualb}
\eeq
A rather obvious procedure would be to follow (\ref{dualb}) by a standard
decimation prescription (star-triangle transformation) in order to get back
to parameters $b'$ defined with respect to the triangular
lattice, and then search for selfdual solutions, $b=b'$.
A key assumption, of course, is that such solutions exist within the
original parameter space, i.e., with only nearest-neighbour couplings
among the spins \cite{Kim74}.

\begin{figure}
\begin{center}
 \leavevmode
 \epsfysize=25mm{\epsffile{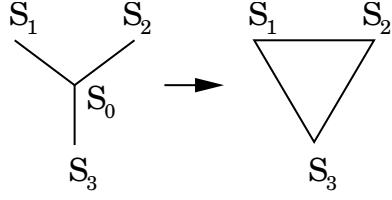}}
 \end{center}
 \protect\caption[3]{The star-triangle transformation.}
\label{fig:decimation}
\end{figure}

The precise setup is shown in Fig.~\ref{fig:decimation}. We form the partial
trace over all spins $S^\mu_0$ situated at even ($Y$-shaped) vertices of the
hexagonal lattice, while keeping the exterior spins $S^\mu_1$, $S^\mu_2$,
$S^\mu_3$ fixed. Defining $\tilde{b}$ as in Eq.~(\ref{defb}), we must have
\beq
 \sum_{S^1_0,S^2_0} \exp \left \lbrace \sum_{i=1}^3 \left(
 \widetilde{K_\mu} \sum_{\mu=1}^2 \delta^\mu_{0i} +
 \widetilde{K_{12}} \delta^1_{0i} \delta^2_{0i} \right) \right \rbrace =
 A \exp \left \lbrace \sum_{i>j=1}^3 \left(
 K'_\mu \sum_{\mu=1}^2 \delta^\mu_{ij} +
 K'_{12} \delta^1_{ij} \delta^2_{ij} \right) \right \rbrace,
\eeq
where the proportionality factor $A$ does not have any bearing on the
duality relations for the coupling constants.

We obtain ten distinct relations by considering all symmetry-unrelated choices
for the fixed spins $S^\mu_i$ with $\mu=1,2$ and $i=1,2,3$. Following
\cite{Kim74} we suppose $q_\mu \ge 3$ integer initially, and then invoke
analytic continuation to claim the validity of the result for arbitrary
$q_\mu$.
\bi
 \item For $S_1 \neq S_2 \neq S_3$ on the two lattices:
% spins all different on the two lattices:
\beq
 (q_1-3)(q_2-3)+3(1+\widetilde{b_1}+\widetilde{b_2}+\widetilde{b_{12}})+
 3(q_2-3)(1+\widetilde{b_1})+3(q_1-3)(1+\widetilde{b_2})+
 6(1+\widetilde{b_1})(1+\widetilde{b_2})=A
 \label{rel1}
\eeq
 \item For $S_1 = S_2 = S_3$ on the two lattices:
% spins identical on the two lattices:
\beq
 (q_1-1)(q_2-1)+(1+\widetilde{b_1}+\widetilde{b_2}+\widetilde{b_{12}})^3+
 (q_2-1)(1+\widetilde{b_1})^3+(q_1-1)(1+\widetilde{b_2})^3=
 A(1+b'_1+b'_2+b'_{12})^3
 \label{rel2}
\eeq
 \item For $S_1 = S_2 \neq S_3$ on the two lattices:
% two spins identical, the same in the two lattices:
\bea
 (q_1-2)(q_2-2)+(1+\widetilde{b_1}+\widetilde{b_2}+\widetilde{b_{12}})^2+
 (1+\widetilde{b_1}+\widetilde{b_2}+\widetilde{b_{12}})+
 (q_2-2)(1+\widetilde{b_1})^2+(q_1-2)(1+\widetilde{b_2})^2+ \nn \\
 (q_2-2)(1+\widetilde{b_1})+(q_1-2)(1+\widetilde{b_2})
 (1+\widetilde{b_1})^2(1+\widetilde{b_2})+
 (1+\widetilde{b_2})^2(1+\widetilde{b_1})=A(1+b'_1+b'_2+b'_{12})
 \label{rel3}
\eea
 \item For $S_1^1 = S_2^1 \neq S_3^1$ and $S_1^2 \neq S_2^2 = S_3^2$:
% two spins identical, not the same in the two lattices:
\bea
 (q_1-2)(q_2-2)+(q_2-2)(1+\widetilde{b_1})^2+
 (q_2-2)(1+\widetilde{b_1})+(q_1-2)(1+\widetilde{b_2})^2+
 (q_1-2)(1+\widetilde{b_2})+ \nn \\
 (1+\widetilde{b_1}+\widetilde{b_2}+\widetilde{b_{12}})
 (1+\widetilde{b_1})(1+\widetilde{b_2})+
 (1+\widetilde{b_1}+\widetilde{b_2}+\widetilde{b_{12}})
 (1+\widetilde{b_1})+(1+\widetilde{b_1})(1+\widetilde{b_2})+ \nn \\
 (1+\widetilde{b_1}+\widetilde{b_2}+\widetilde{b_{12}})(1+\widetilde{b_2})=
 A(1+b'_1)(1+b'_2)
 \label{rel4}
\eea
\item For $S_1^1 = S_2^1 = S_3^1$ and $S_1^2 \neq S_2^2 \neq S_3^2$:
% spins identical in the first lattice, different in the second: 
\beq
 (q_1-1)(q_2-3)+(q_2-3)(1+\widetilde{b_1})^3+3(q_1-1)(1+\widetilde{b_2})+
 3(1+\widetilde{b_1}+\widetilde{b_2}+\widetilde{b_{12}})(1+\widetilde{b_1})^2=
 A(1+b'_1)^3
 \label{rel5}
\eeq
%\item For $S_1^1 \neq S_2^1 \neq S_3^1$ and $S_1^2 = S_2^2 = S_3^2$:
%% spins different in the first, identical in the second:
%\beq
% (q_1-3)(q_2-1)+(q_1-3)(1+\widetilde{b_2})^3+3(q_2-1)(1+\widetilde{b_1})+
% 3(1+\widetilde{b_1}+\widetilde{b_2}+\widetilde{b_{12}})(1+\widetilde{b_2})^2=
% A(1+b'_2)^3
% \label{rel6}
%\eeq
%\item For $S_1^1 = S_2^1 \neq S_3^1$ and $S_1^2 = S_2^2 = S_3^2$:
%% two spins identical in the first, all identical in the second:
%\bea
% (q_1-2)(q_2-1)+(q_2-1)(1+\widetilde{b_1})^2+(q_2-1)(1+\widetilde{b_1})+
% (q_1-2)(1+\widetilde{b_2})^3+ \nn \\
% (1+\widetilde{b_1}+\widetilde{b_2}+
% \widetilde{b_{12}})^2(1+\widetilde{b_2})+
% (1+\widetilde{b_1}+\widetilde{b_2}+\widetilde{b_{12}})(1+\widetilde{b_2})^2=
% A(1+b'_1+b'_2+b'_{12})(1+b'_2)^2
% \label{rel7}
%\eea
\item For $S_1^1 = S_2^1 = S_3^1$ and $S_1^2 = S_2^2 \neq S_3^2$:
% spins identical in the first, two identical in the second:
\bea
 (q_2-2)(q_1-1)+(q_1-1)(1+\widetilde{b_2})^2+(q_1-1)(1+\widetilde{b_2})+
 (q_2-2)(1+\widetilde{b_1})^3+ \nn \\
 (1+\widetilde{b_1}+\widetilde{b_2}+
 \widetilde{b_{12}})^2(1+\widetilde{b_1})+(1+\widetilde{b_1}+\widetilde{b_2}+
 \widetilde{b_{12}})(1+\widetilde{b_1})^2=
 A(1+b'_1+b'_2+b'_{12})(1+b'_1)^2
 \label{rel8}
\eea
\item For $S_1^1 = S_2^1 \neq S_3^1$ and $S_1^2 \neq S_2^2 \neq S_3^2$:
% two identical in the first, different in the second:
\bea
 (q_1-2)(q_2-3)+(q_2-3)(1+\widetilde{b_1})^2+(q_2-3)(1+\widetilde{b_1})+
 (q_1-2)3(1+\widetilde{b_2})+ \nn \\
 2(1+\widetilde{b_1}+\widetilde{b_2}+
 \widetilde{b_{12}})(1+\widetilde{b_1})+
 (1+\widetilde{b_1})^2
 (1+\widetilde{b_2})+(1+\widetilde{b_1}+\widetilde{b_2}+\widetilde{b_{12}})+
 2(1+\widetilde{b_1})(1+\widetilde{b_2})=A(1+b'_1)
 \label{rel9}
\eea
%\item For $S_1^1 \neq S_2^1 \neq S_3^1$ and $S_1^2 = S_2^2 \neq S_3^2$:
%% spins different in the first, two identical in the second:
%\bea
% (q_2-2)(q_1-3)+(q_1-3)(1+\widetilde{b_2})^2+(q_1-3)(1+\widetilde{b_2})+
% (q_2-2)3(1+\widetilde{b_1})+ \nn \\
% 2(1+\widetilde{b_1}+\widetilde{b_2}+
% \widetilde{b_{12}})(1+\widetilde{b_2})+
% (1+\widetilde{b_2})^2
% (1+\widetilde{b_1})+(1+\widetilde{b_1}+\widetilde{b_2}+
% \widetilde{b_{12}})+2(1+\widetilde{b_2})(1+\widetilde{b_1})=A(1+b'_2)
% \label{rel10}
%\eea
\item Eqs.~(\ref{rel5}), (\ref{rel8}) and (\ref{rel9}) each represent a pair
 of relations of which we have only written one representative; the other
 one is obtained by permuting the two models, i.e., by letting
 $q_1 \leftrightarrow q_2$ and $b_1 \leftrightarrow b_2$.
\ei

\subsection{General structure of trivial solutions}

The list of selfdual solutions of $N$ coupled models, $\mu=1,2,\ldots,N$, will
in general contain a certain number of trivial solutions. By a trivial
solution we mean one which is a consequence of the selfduality of a single
model. Let us discuss in detail two classes \cite{Jacobsen00,Dotsenko02}
of trivial solutions:
\be
 \item The models are actually decoupled. This happens, e.g., in the above
 example with $N=2$ when $K_{12}=0$, that is $b_{12} = b_1 b_2$. We will then
 have $b_\mu^3 + 3 b_\mu^2 = q_\mu$ for each $\mu$, cf.~Eq.~(\ref{trisol}).
 The number of real solutions of the $\mu$'th equation is $n_\mu = 3$ when
 $0 < q_\mu < 4$, $n_\mu = 2$ when $q_\mu=0$ or $q_\mu=4$, and $n_\mu = 1$
 otherwise; there will therefore be $n=\prod_{\mu=1}^N n_\mu$ trivial
 solutions for the system of $N$ decoupled models.
 \item The models couple strongly so as to form a {\em single}
 $q=\prod_{\mu=1}^N q_\mu$ state model. This happens when only the coupling
 constant involving all the models is nonzero. E.g., in the above example with
 $N=2$, one would have $K_1=K_2=0$, that is $b_1=b_2=0$. The number of such
 solutions equals the number of real solutions of Eq.~(\ref{trisol}), with $b$
 replaced by $b_{12}$.
\ee
The goal of our study is to show that there exists selfdual solutions of
coupled Potts models on the triangular lattice which are {\em not} trivial in
the above sense.
 
\subsection{Non-trivial solutions}
\label{sec:N2L0}

Let us return to the Hamiltonian (\ref{H2}). We have numerically solved
the ten relations (\ref{rel1})--(\ref{rel9}) for several different values of
$q_1$ and $q_2$. The conclusion is that for $q_1 \neq q_2$ only trivial
solutions exist.

For $q_1=q_2$ the situation is different. There are now only seven distinct
relations (\ref{rel1})--(\ref{rel9}), since the three relations which formerly
occurred in pairs will now collapse into single relations. The parameters
$b$ are thus less constrained, and accordingly we find non-trivial solutions.
Numerically we find that these solutions have $b_1=b_2$ (but note that there
are still trivial solutions with $K_{12}=0$ which break this symmetry).

%These
%solutions are given in Fig.~\ref{table} for the first integer values of $q$.

%\begin{figure}
%\begin{center}
%\begin{tabular} {|c|c|c|c|c|c|}
%\hline $~~~~g~~~~$ & $~~~~q~~~~$ & $~~~b_1~~~$ & $~~~b_{12}~~~$ & $~~~e^{K_1}~~~$ & $~~~e^{K_{12}}~~~$ \\
%\hline 0 & 4 & -2 & 1 & -1 & -2 \\
%\hline $\frac{1}{6}$ & 3 & -1.653 & 0.773 & -0.653 & -3.596 \\
%\hline $\frac{1}{4}$ & 2 & -1.268 & 0.536 & -0.268 & -13.928 \\
%\hline $\frac{1}{3}$ & 1 & -0.815 & 0.283 & 0.185 & -10.170 \\
%\hline $\frac{1}{2}$ & 0 & 0 & 0 & 1 & 1 \\
%\hline $\frac{2}{3}$ & 1 & 0.227 & 0.426 & 1.227 & 1.249 \\
%\hline $\frac{3}{4}$ & 2 & 0 & 1 & 1 & 2 \\
%\hline $\frac{5}{6}$ & 3 & -0.468 & 1.815 & 0.532 & 6.638 \\
%\hline 1 & 4 & -2 & 4 & -1 & 1 \\
%\hline $\frac{7}{6}$ & 3 & -3.879 & 6.411 & -2.879 & -0.042 \\
%\hline $\frac{5}{4}$ & 2 & -4.732 & 7.464 & -3.732 & -0.072 \\
%\hline $\frac{4}{3}$ & 1 & -5.411 & 8.291 & -4.411 & -0.079 \\
%\hline $\frac{3}{2}$ & 0 & -6 & 9 & -5 & $-\frac{2}{25}$ \\
%\hline
%\end{tabular} 
%\end{center}
%\caption{Non-trivial solutions for $q$ integer.}
%\label{table}
%\end{figure}

Setting now $q \equiv q_1=q_2$ and $b_1=b_2$ we can obtain the
non-trivial solutions analytically, e.g., by solving Eqs.~(\ref{rel1}),
(\ref{rel3}) and
(\ref{rel9}) for $A$, $b_1$ and $b_{12}$, and verifying that the found
solution satisfies all the other relations. The result is:
\beq
 b_1^3 + 6b_1^2 + 3b_1 q+ q(q-2)=0, \qquad
 b_{12}=\frac{q-b_1^2}{2+b_1}.  \label{N2sol}
\eeq

\begin{figure}
\begin{center}
 \leavevmode
 \epsfysize=60mm{\epsffile{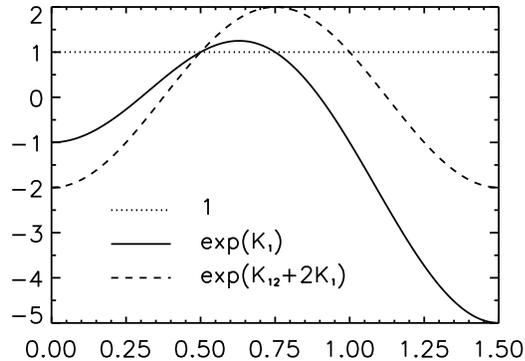}}
 \end{center}
 \protect\caption[3]{Variation of the Boltzmann weights (\ref{Boltzmann})
with the parameter $g$ defined in Eq.~(\ref{N2sola}).}
\label{fig:couplings}
\end{figure}

For each $q \in (0,4)$ Eq.~(\ref{N2sol}) admits three distinct solution for
$b_1$. To make clear in the following exactly to which solution we are
referring, it is convenient to recast (\ref{N2sol}) in parametric form,
by setting $q=4\cos^2(\pi g)$. When the parameter $g$ runs through the
interval $[0,\frac32]$, the number of states $q$ runs through the interval
$[0,4]$ three times. We have then
\beq
 b_1    = x(1-x), \qquad
 b_{12} = (x-1)^2, \qquad
 x \equiv 2 \cos \left(\frac{2\pi}{3} g \right). \label{N2sola}
\eeq
This parametrisation also has the advantage over (\ref{N2sol}) that it is
non-singular as $g \to 1$ (i.e., $x\to -1$ and $q \to 4$) and yields the
correct limiting values of $b_1$ and
$b_{12}$.

In terms of $x$, the Boltzmann weights for two neighbouring spins
being identical in none, one, or both of the two models read:
\beq
 1, \qquad
 {\rm e}^{K_1} = 1+x-x^2, \qquad
 {\rm e}^{K_{12}+2 K_1} = 2 - x^2. \label{Boltzmann}
\eeq
Their variation with $g$ is shown in Fig.~\ref{fig:couplings}. Note that
the $K_{12}$ coupling is physical (${\rm e}^{K_{12}} \ge 0$) for
$\frac38 \le g \le \frac98$, and that $K_1$ is physical for
$\frac{3}{10} \le g \le \frac{9}{10}$.

It is also interesting to remark that in the $x$-parametrisation,
Eq.~(\ref{trisol}) for a single model reads $b=x-1$; the trivial solution of
type 1 is then $b_1 = x-1$, $b_{12}=(b_1)^2$ with the {\em same} value of
$b_{12}$ as in Eq.~(\ref{N2sola}).

\subsection{Special points on the curve (\ref{N2sola})}
\label{sec2d}

Let us remark on a few special values of the parameter $g$ for which the
physics of the two coupled models can be related to that of a single model.
\be
 \item For $g=1$ (i.e., $q=4$ and $x=-1$) one has $K_{12}=0$,
 whence the models are decoupled.
 \item For $g=\frac34$ (i.e., $q=2$ and $x=0$) one has $K_1=0$ and
 $K_{12}=\log 2$. Thus, the two models couple strongly to form a single
 $q^2=4$ state model at the ferromagnetic fixed point.
 \item For $g=\frac12$ (i.e., $q=0$ and $x=1$) one has $K_1=K_{12}=0$.
 This is an infinite-temperature limit, whose properties depend on the
 ratio $K_1/K_{12}$ as $g\to\frac12$. In fact, for $x \to 1$ we find
 $K_1 \sim (1-x)$ and $K_{12} \sim 2(1-x)^3$, whence the coupling between
 the two models is negligible. Note also that $q \sim 3(1-x)^2$, whence
 the point $(q,b_1)=(0,0)$ is approached with infinite slope, as is the case
 for a single Potts model {\em along} the selfdual curve (\ref{trisol}).
\ee
Note also that these values of $g$ correspond to
degeneracies of the Boltzmann weights (\ref{Boltzmann}),
cf.~Fig.~\ref{fig:couplings}.

As to the critical behaviour of a single Potts model, the situation is
well understood in the case of the square lattice \cite{Saleur91}. There
are three critical phases, referred to as ferromagnetic, Berker-Kadanoff
and antiferromagnetic in Ref.~\cite{Saleur91}. By universality, one would
expect the same three critical phases to describe the distinct branches
of the selfdual curve (\ref{trisol}), as is confirmed by numerical transfer
matrix results \cite{Jacobsen04}. In particular, for the central charge
along (\ref{trisol}) one has
\bea
 c &=& 1-\frac{6 g^2}{1-g}, \quad \qquad \mbox{for } 0 \le g < \frac12
 \qquad \mbox{(ferromagnetic phase)} \nn \\
 c &=& 1-\frac{6 g^2}{1-g}, \quad \qquad \mbox{for } \frac12 \le g < 1
 \qquad \mbox{(Berker-Kadanoff phase)} \nn \\
 c &=& 2-6(g-1), \qquad \mbox{for } 1 < g \le \frac32
 \qquad \mbox{(antiferromagnetic phase)} \label{c1model}
\eea
when parametrising $q=4 \cos^2(\pi g)$ as in (\ref{N2sola}).
Note that points $(q,b_1)=(0,0)$ and $(q,b_1)=(4,-2)$ are special, and the
critical behaviour when approaching the curve (\ref{trisol}) at these points
depends on the exact prescription for taking the limit.

We conclude that the above-mentioned special points on the selfdual curve
(\ref{N2sola}) for two coupled models lead to the central charges
$c=-4$ for $g=\frac12$ and $c=1$ for $g=\frac34$. For $g\to 1$, the result
is for the moment uncertain, due to the ambiguity in taking the limit
just referred to.
%and $c=4$ for $g=1$.

\section{Two and three-spin interactions}
\label{sec3}

The results of the preceding section can be generalised to a model defined
by the Hamiltonian $H=H_2+H_3$. The term $H_2$ is as in Eq.~(\ref{H2}),
whereas $H_3$ introduces interactions between the three spins around the
up-pointing faces $\langle ijk \rangle$ of the triangular lattice,
as shown in Fig.~\ref{fig:lattice}:
\beq
 \beta H_3 = -\sum_{\langle ijk \rangle} \left \lbrace
 L_1 \delta^1_{ijk} + L_2 \delta^2_{ijk} +
 L_{12} \delta^1_{ijk} \delta^2_{ijk} \right \rbrace.
\eeq
Introducing such interactions around {\em every} face seems difficult,
even in the case of a single model \cite{Wu82}.

One could of course consider also interactions between two spins in one model
and three spins in the other. It can be verified that the method given below
can be adapted to this case. However, we have chosen not to consider any
further such mixed interactions.

In the following it is convenient to introduce the parameters
\beq
 y_\mu  = {\rm e}^{L_\mu} - 1, \qquad
 y_{12} = {\rm e}^{L_{12}+L_1+L_2} -
 {\rm e}^{L_1} - {\rm e}^{L_2} + 1 \label{defy}
\eeq
in analogy with Eq.~(\ref{defb}).
Guided by our results without three-spin interactions, we shall assume
in the following that non-trivial selfdual solutions only exist when
coupling identical models. Thus we restrict the study to $q \equiv q_1 = q_2$,
$b_1 = b_2$ and $L_1 = L_2$.
% (We have however not tested this numerically.)

\subsection{Mapping to a fully-packed loop model}

Wu and Lin \cite{WuLin80} have shown how to produce duality relations for
a single Potts model with two- and three-spin interactions, by mapping it
to a related loop model. After briefly reviewing their method, we shall
adapt it to the case of coupled models.

\begin{figure}
\begin{center}
 \leavevmode
 \epsfysize=25mm{\epsffile{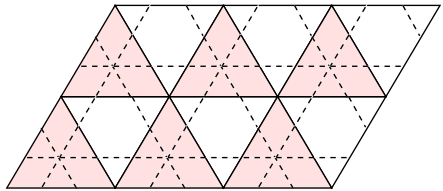}}
 \end{center}
 \protect\caption[3]{The lattice of Potts spins is shown in solid linestyle.
There are two-spin interactions along every edge, and three-spin interactions
among the spins surrounding the up-pointing faces (shaded). The loop model is
defined on a shifted triangular lattice, shown in dashed linestyle.}
\label{fig:lattice}
\end{figure}

\begin{figure}
\begin{center}
 \leavevmode
 \epsfysize=35mm{\epsffile{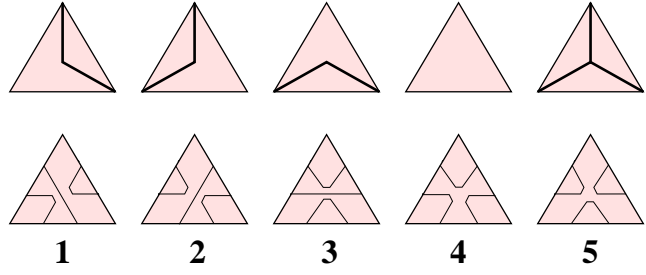}}
 \end{center}
 \protect\caption[3]{Correspondence between link diagrams ${\cal L}$
for the Potts model (first line) and vertices ${\cal L}'$ of the fully-packed
loop model (second line).}
\label{fig:mapping}
\end{figure}

In Fig.~\ref{fig:lattice} we show the triangular lattice of Potts spins,
and the shifted triangular lattice on which the loop model is defined.
To obtain the correspondence, one first rewrites the Boltzmann weight around
an up-pointing triangle $\langle ijk \rangle$ as
\beq
 w_{ijk} = f_1 \delta_{ij} + f_2 \delta_{jk} + f_3 \delta_{ik} +
 f_4  + f_5 \delta_{ijk} \equiv \sum_{a=1}^5 f_a \delta_a, \label{5terms}
\eeq
where $\delta_1 \equiv \delta_{ij} = \delta(S_i,S_j)$ etc. Note that $f_4 = 1$.
To each of the five terms in this
sum is associated a link diagram on $\langle ijk \rangle$, as shown in the
first line of Fig.~\ref{fig:mapping}, indicating which spins participate in the
delta symbol. The partition function is then
\beq
 Z_{\rm Potts}=\sum_{S_i} \prod_{\langle ijk \rangle} w_{ijk}
  =\sum_{\cal L} q^{n({\cal L})} \prod_{a=1}^5 f_a^{n_a({\cal L})},
\eeq
where the sum is over all link diagrams ${\cal L}$ for the whole lattice,
$n({\cal L})$ is the number of connected components in ${\cal L}$, and 
$n_a({\cal L})$ is the number of up-pointing triangles whose link diagram
is of type $a=1,2,\ldots,5$.

The link diagrams are now mapped to fully-packed loop configurations
on a shifted triangular lattice (cf.~Fig.~\ref{fig:lattice}) via the
correspondence given in the second line of Fig.~\ref{fig:mapping}.
The partition function of the loop model is defined as
\beq
 Z_{\rm loop}=\sum_{\cal L'} z^{p({\cal L'})}
 \prod_{a=1}^5 c_a^{n_a({\cal L'})},
\eeq
where $p({\cal L'})$ is the number of closed polygons (loops) in the loop
configuration ${\cal L}'$ (which is in one-to-one correspondence with ${\cal
L}$ using Fig.~\ref{fig:mapping}), and $z$ is the fugacity of a polygon.
Using now the Euler relation one finds that
$Z_{\rm Potts} = Z_{\rm loop}$ if $z = \sqrt{q}$, $c_a = f_a$ for $a=1,2,3$,
$c_4 = q^{1/2} f_4$, and $c_5 = q^{-1/2} f_5$ \cite{WuLin80}.

Finally, the duality relation follows from the invariance of the loop model
under $\pi/3$ rotations; notice that this cyclically interchanges the link
diagrams of types 1, 2, 3 and permutes the diagrams of type 4, 5.

\subsection{$N=2$ and mapping to coupled loop models}

The above mapping can be adapted to the case of $N$ coupled models. We here
consider $N=2$. The Boltzmann weight $w_{ijk}$ around an up-pointing triangle
$\langle ijk \rangle$ can be written in a form generalising (\ref{5terms}):
\beq
 w_{ijk} = \sum_{a,b=1}^5 f_{ab} \delta_a^1 \delta_b^2 \, ;
\eeq
as usual $\delta^\mu$ refers to the $\mu$'th model. As the two models are
identical, we have the symmetry $f_{ab} = f_{ba}$.
When $L_1=L_2=L_{12}=0$, the couplings are denoted $f^{0}_{ab}$; they are
related to the $b$ through
\bea
 f^{0}_{ii} &=& b_{12}, \nn \\
 f^{0}_{ij} &=& b_1^2, \nn \\
 f^{0}_{i4} &=& b_1, \nn \\
 f^{0}_{i5} &=& b_{12} (b_1^2+2 b_1)+b_1^3, \nn \\
 f^{0}_{44} &=&1, \nn \\
 f^{0}_{45} &=& b_1^3+3 b_1^2, \nn \\
 f^{0}_{55} &=& b_{12}^3+3 b_{12}^2(1+2 b_1)+6 b_{12} b_1^2,
\eea
where $i \neq j$ are any numbers in $\{1,2,3\}$.
In the general case ($L \neq 0$) we then have
\bea
 f_{14} &=& f_{14}^0, \nn \\
 f_{15} &=& f_{15}^0+y_1(f_{11}^0+2f_{12}^0+f_{14}^0+f_{15}^0), \nn \\
 f_{44} &=& f_{44}^0, \nn \\
 f_{55} &=& f_{55}^0+2 y_1 (3 f_{15}^0+f_{45}^0+f_{55}^0)+
 y_{12} (3 f_{11}^0+6 f_{12}^0+6 f_{14}^0+6 f_{15}^0+f_{44}^0+
 2 f_{45}^0+f_{55}^0) \, ;
\eea
note that we here only give the $f_{ab}$ needed in the duality relation
(\ref{tworels}) below.

The Potts model partition function reads
\beq
 Z_{\rm Potts} =\sum_{{\cal L}_1,{\cal L}_2} q^{n({\cal L}_1)+n({\cal L}_2)}
 \prod_{a,b=1}^5 f_{ab}^{n_{ab}({\cal L}_1,{\cal L}_2)}. \label{N2Potts}
\eeq
Here, $n_{ab}({\cal L}_1,{\cal L}_2)$ is the number of up-triangles where
model 1 is in the link state $a$ and model 2 is in the link state $b$.
On the other hand, we can define a coupled loop model through
\beq
 Z_{\rm loop}=\sum_{{\cal L}'_1,{\cal L}'_2} z^{p({\cal L}'_1)+p({\cal L}'_2)}
 \prod_{a,b=1}^5 c_{ab}^{n_{ab}({\cal L}'_1,{\cal L}'_2)}. \label{N2Loop}
\eeq
The equivalence between the two goes through as before. Using the Euler
relation, one finds $Z_{\rm Potts}=Z_{\rm loop}$ provided that
$z=\sqrt{q}$ and that
\beq
 c_{ab} = q^{(\delta_{a,4}+\delta_{b,4}-\delta_{a,5}-\delta_{b,5})/2}
 f_{ab}. \label{N2equiv}
\eeq
It should now be obvious how the mapping generalises to an arbitrary number
of coupled models. One has $c_{ab\cdots}=q^{(N_4-N_5)/2} f_{ab\cdots}$,
where $N_4$ (resp.\ $N_5$) counts the number of indices equal to 4 (resp.\ 5).

The selfduality criterion is again obtained by requiring invariance of
$Z_{\rm loop}$ under $\pi/3$ rotations. This means that the $c_{ab}$ are
invariant under independent permutations of the indices $\{1,2,3\}$ and
of $\{4,5\}$. We also recall the invariance under a permutation of the two
models, $c_{ab}=c_{ba}$. Actually, since the two Potts model were taken to
be identical and isotropic from the outset, the only non-trivial selfduality
criteria are:
\beq
 c_{44}=c_{55}, \qquad
 c_{14}=c_{15}. \label{tworels}
\eeq

\subsection{Selfdual solutions}

We now wish to express the condition of selfduality in terms of the parameters
$b$ and $y$, cf.~Eqs.~(\ref{defb}) and (\ref{defy}).

When three-spin interactions are absent ($L=0$), the relations (\ref{tworels})
immediately yield the solutions given in Sec.~\ref{sec:N2L0}.
This is a remarkable simplification when compared to solving the system of
relations (\ref{rel1})--(\ref{rel9}); indeed many of these relations turn
out to be dependent.

When $L \neq 0$, Eq.~(\ref{tworels}) still gives the complete solution to
the selfduality problem, but it does not generically lead to simple
expressions in terms of the parameters $b$ and $y$. We therefore concentrate
on a few remarkable solutions. As before, there are two types of trivial
solutions:
\be
 \item Trivial decoupled solution ($K_{12}=L_{12}=0$, or $b_{12}=b_1^2$ and
 $y_{12}=y_1^2$). One finds the selfduality criterion of a {\em single} model
 \cite{Baxter78,WuLin80}
 \beq
  b_1^3+3\,b_1^2-q+y_1(1+b_1)^3=0. \label{sd1model}
 \eeq
 \item Trivial strongly coupled solution ($K_1=L_1=0$, or $b_1=y_1=0$).
 We find
 \beq
  b_{12}^3+3 b_{12}^2-q^2+y_{12}(1+b_{12})^3=0.
 \eeq
 This is just the selfduality criterion of a
 single $q^2$ state model.
\ee
Some noteworthy non-trivial solutions can be found by giving particular values
to $b_1$, $y_1$ or $y_{12}$:
\be
 \item
 For $y_1=0$ (i.e., $L_1=0$), there is only one non-trivial solution:
% which extends (\ref{N2sol}) to arbitrary $y_{12}$:
 \beq
%   b_1^6+9 b_1^5+(3 q+18)b_1^4+(q^2+6 q)b_1^3+
%   (3 q^2-12 q)b_1^2-3 q^2 b_1-q^3+2 q^2=y_{12} (q+5 b_1+2+b1^2)^3, \nn \\
   \left( b_1^3 + 6 b_1^2 + 3 q b_1 + q(q-2) \right)
   \left( b_1^3 + 3 b_1^2 - q \right) =y_{12} (b_1^2+5 b_1+q+2)^3, \qquad
   b_{12}=\frac{q-b_1^2}{2+b_1}.
 \eeq
 Note that when $y_{12}=0$, the factorisation of the left-hand side allows
 us to retrieve either (\ref{N2sol}) or the trivial solution (\ref{sd1model}).
 \item For $y_{12}=y_1^2$ (i.e., $L_{12}=0$), there is a solution of the form:
 \beq
   b_1=-\frac{q}{2}, \qquad
   b_{12}=-\frac{q^2}{2}+3 q-3, \qquad
   y_1=\frac{q(4-q)}{(q-2)^2}.
 \eeq
 % There is also another complicated solution.
 \item For $b_1=-1$ (i.e., $K_1 \to -\infty$), there is a solution of the form:
 \beq
   b_{12} = q-1, \qquad
   y_1 =  -\frac12 \left( y_{12} + \frac{q^2-5q+5}{q^2-4q+4} \right).
 \eeq
% \item For $y_{12}=0$, there is a solution of the form:
% \bea
%  (2 q^2-8 q-8)b_{12}^3+(30 q^2-6 q^3-24)b_{12}^2+(36 q^2-36 q^3+6 q^4)
%  b_{12}+8 q^2-20 q^3+13 q^4-2 q^5=0, \nn \\
%  b_1=-\frac{q}{2}, \qquad y_1=\frac{q(4-q)}{(q-2)^2}.
% \eea 
% There is also another complicated solution.
% Note that these solutions correspond to complex values of $L_{12}$.
\ee

\section{Three coupled models}
\label{sec4}

The technique of mapping to coupled loop models has permitted us to study
the case of $N=3$ coupled Potts models, defined by the Hamiltonian
\bea
 \beta H &=& -\sum_{\langle ij \rangle} \left \lbrace
 K_1 \sum_{\mu=1}^3 \delta^\mu_{ij} +
 K_{12} \sum_{\mu > \nu = 1}^3 \delta^\mu_{ij} \delta^\nu_{ij} +
 K_{123} \delta^1_{ij} \delta^2_{ij} \delta^3_{ij} \right \rbrace \nn \\
 & & -\sum_{\langle ijk \rangle} \left \lbrace
 L_1 \sum_{\mu=1}^3 \delta^\mu_{ijk} +
 L_{12} \sum_{\mu > \nu = 1}^3 \delta^\mu_{ijk} \delta^\nu_{ijk} +
 L_{123} \delta^1_{ijk} \delta^2_{ijk} \delta^3_{ijk} \right \rbrace.
\eea
Since in the case of two coupled models, nontrivial selfdual solutions were
only found when coupling identical models in an isotropic way, we shall
henceforth restrict the coupling constants as follows:
\beq
 K_1=K_2=K_3, \qquad
 K_{12}=K_{13}=K_{23}, \qquad
 L_1=L_2=L_3, \qquad
 L_{12}=L_{13}=L_{23},
\eeq
and we will use the parameters \cite{Jacobsen00,Dotsenko02}
\bea
 b_1={\rm e}^{K_1}-1, \qquad
 b_{12}={\rm e}^{K_{12}+2K_1}-2{\rm e}^{K_1}+1, \qquad
 b_{123}={\rm e}^{K_{123}+3K_{12}+3K_1}-3{\rm e}^{K_{12}+2K_1}+3{\rm e}^{K_1}-1 \\
 y_1={\rm e}^{L_1}-1, \qquad
 y_{12}={\rm e}^{L_{12}+2L_1}-2{\rm e}^{L_1}+1, \qquad
 y_{123}={\rm e}^{L_{123}+3L_{12}+3L_1}-3{\rm e}^{L_{12}+2L_1}+3{\rm e}^{L_1}-1.
\eea

The mapping to coupled loop models follows the obvious generalisation
of Eqs.~(\ref{N2Potts})--(\ref{N2Loop}), and the equivalence criterion
is stated in the remark after Eq.~(\ref{N2equiv}).

To relate the coupling constants $f_{abc}$ to the $b$ and $y$,
we first consider the case of vanishing three-spin interactions
(i.e., $y=y_{12}=y_{123}=0$). Letting $i \neq j \neq k$ designate distinct
numbers in $\{1,2,3\}$ we have:
\bea
 f_{iii}^0 &=& b_{123}, \nn \\
 f_{iij}^0 &=& b_1 b_{12}, \nn \\
 f_{ijk}^0 &=& b_1^3, \nn \\
 f_{ii4}^0 &=& b_{12}, \nn \\
 f_{ii5}^0 &=& b_{123} b_1^2+2 b_{123} b_1+b_{12} b_1^2, \nn \\
 f_{ij4}^0 &=& b_1^2, \nn \\
 f_{ij5}^0 &=& b_{12}^2 (1+b_1)+2 b_{12} b_1^2, \\
 f_{i44}^0 &=& b_1, \nn \\
 f_{i45}^0 &=& b_{12} (b_1^2 + 2 b_1) + b_1^3, \nn \\
 f_{i55}^0 &=& 2 b_{12}^3+5 b_{12}^2 b_1+2 b_{123} b_{12}+
   4 b_{123} b_{12} b_1+2 b_{123}b_1^2+b_{123} b_{12}^2, \nn \\
 f_{444}^0 &=& 1, \nn \\
 f_{445}^0 &=& b_1^3+3 b_1^2, \nn \\
 f_{455}^0 &=& b_{12}^3+3 b_{12}^2+6 b_{12} b_1^2+6 b_{12}^2 b_1, \nn \\
 f_{555}^0 &=& 6 b_{12}^3+18 b_{12} (b_1+b_{12}) b_{123}+
     3 (1+3 b_1+3 b_{12}) b_{123}^2+b_{123}^3. \nn
\eea
For the general case one then has (note that we only give those $f_{abc}$
which will be used in the duality relation (\ref{threerel}) below):
\bea
 f_{114} &=& f_{114}^0, \nn \\
 f_{115} &=& f_{115}^0+y_1 (f_{111}^0+2 f_{112}^0+f_{114}^0+f_{115}^0), \nn \\
 f_{124} &=& f_{124}^0, \nn \\
 f_{125} &=& f_{125}^0+y_1 (2 f_{112}^0+f_{123}^0+f_{124}^0+f_{125}^0), \nn \\
 f_{144} &=& f_{144}^0, \nn \\
 f_{155} &=& f_{155}^0+2 y (f_{115}^0+2 f_{125}^0+f_{145}^0+f_{155}^0)+y_{12} (f_{111}^0+f_{144}^0+f_{155}^0+6 f_{112}^0+2 f_{123}^0+ \\
 &&2 f_{114}^0+2 f_{115}^0+4 f_{124}^0+4 f_{125}^0+2 f_{145}^0), \nn \\
 f_{444} &=& f_{444}^0, \nn \\
 f_{445} &=& f_{445}^0+y (3 f_{144}^0+f_{444}^0+f_{445}^0), \nn \\
 f_{455} &=& f_{455}^0+2 y (3 f_{145}^0+f_{445}^0+f_{455}^0)+
   y_{12} (3 f_{114}^0+f_{444}^0+f_{455}^0+6 f_{124}^0+6 f_{144}^0+
   6 f_{145}^0+2 f_{445}^0), \nn \\
 f_{555} &=& f_{555}^0+3 y (3 f_{155}^0+f_{455}^0+f_{555}^0)+
   3 y_{12} (3 f_{115}^0+f_{445}^0+f_{555}^0+6 f_{125}^0+
   6 f_{145}^0+6 f_{155}^0+2 f_{455}^0)+y_{123} (3 f_{111}^0+ \nn \\
   &&f_{444}^0+f_{555}^0+18 f_{112}^0+9 f_{114}^0+9 f_{115}^0+
   6 f_{123}^0+ 18 f_{124}^0+18 f_{125}^0+9 f_{144}^0+9 f_{155}^0+
   18 f_{145}^0+3 f_{445}^0+3 f_{455}^0). \nn
\eea

So from (the generalisation of) Eq.~(\ref{N2equiv}) we know how to express
the $c_{abc}$ in terms of the $b$ and $y$. Given that we have 
isotropic Potts models with the same coupling constants, the non-trivial selfduality relations
read simply:
\beq
 c_{444}=c_{555}, \qquad
 c_{455}=c_{445}, \qquad
 c_{155}=c_{144}, \qquad
 c_{115}=c_{114}, \qquad
 c_{125}=c_{124}. \label{threerel}
\eeq

\subsection{Selfdual solutions}

There are two types of trivial solutions, as in the case of two coupled models.

An important difference with the case of two coupled models is that
non-trivial selfdual solutions with $y=y_{12}=y_{123}=0$ only exist for
exceptional values of $q$ (i.e., $q=0$ and $q=4$; see below). So for generic
values of $q$, the three-spin interactions are necessary to generate
non-trivial solutions of the selfduality problem.

As before, the most general solutions are not algebraically simple in terms
of the variables $b$ and $y$. We therefore report only a few special cases:
\be
 \item There are three non-trivial solutions with $L_{12}=L_{123}=0$ (i.e.,
 $y_{12}=y_1^2$ and $y_{123}=y_1^3$). They read:
 \bea
 && b_1 = -\frac{q}{2}, \qquad
 b_{12}=\frac{q^2}{4}, \qquad
 b_{123}=\frac{q^3-9 q^2+18 q-12}{4}, \qquad
 y_1=\frac{q (4-q)}{q^2-4 q+4} \, ; \\
 && b_1 = -1, \qquad
 b_{12}=\frac{q}{2}, \qquad
 b_{123}=\frac{q (1-q)}{2}, \qquad
 y_1=\frac{q}{2-q} \, ; \\
 && (4 q-6) b_1^2+2 q b_1+q=0, \qquad
 b_{12}=-\frac{q(1+2b_1)}{2}, \nn \\
 && \quad
 b_{123}=\frac{q^2 ((8 q^2-16 q-6) b_1+4 q^2-12 q+3)}{4 (3 b_1+q) (2 q-3)},
 \qquad
 y_1=\frac{q}{2-q}.
 \eea
 \item There are two non-trivial solutions with $L_1=0$ (i.e., $y_1=0$):
 \bea
 && b_1=-1, \qquad
 b_{12}=\frac{q}{2}, \qquad
 b_{123}=\frac{q (1-q)}{2}, \qquad
 y_{12}=\frac{q (4-q)}{q^2-4 q+4}, \qquad
 y_{123}=\frac{2 q (q^2-6 q+6)}{q^3-6 q^2+12 q-8} \, ; \\
 && b_1=1-q, \qquad
 b_{12}=\frac{q (1-q)}{2-q}, \qquad
 b_{123}=\frac{q (q^2-3 q+1)}{(2-q)(q-3)}, \nn \\
 && \quad
 y_{12}=\frac{q (q^6-11 q^5+45 q^4-87 q^3+86 q^2-42 q+8)}
 {q^6-18 q^5+126 q^4-432 q^3+756 q^2-648 q+216}, \nn \\
 && \quad y_{123}=\frac{q (4 q^8-72 q^7+531 q^6-2068 q^5+4584 q^4-5856 q^3+4220 q^2-1584 q+240)}{q^9-30 q^8+390 q^7-2872 q^6+13140 q^5-38520 q^4+71928 q^3-82080 q^2+51840 q-13824}.
 \eea
\item There are two non-trivial solutions with $y_{12}=0$. We give here only
 the first one because the other is complicated:
 \beq
 b_1=-1, \qquad
 b_{12}=\frac{q}{2}, \qquad
 b_{123}=\frac{q(1-q)}{2}, \qquad % (and therefore $b_{123}(q+1)=b_{123}(q)-q$)
 y_1=\frac{q (4-q)}{2 (q^2-4 q+4)}, \qquad
 y_{123}=\frac{q^2 (q-6)}{2 (q^3-6 q^2+12 q-8)}.
 \eeq
\ee
Note that for $q=2$ the non-trivial solutions given are singular. In fact,
for $q=2$, they can be written as: 
$b_1=-1$, $b_{12}=1$, $b_{123}=-1$, and the values of
$y$, $y_{12}$, $y_{123}$ are arbitrary. 
Indeed the values of the $b$ correspond to three decoupled antiferromagnetic
Potts models at zero temperature, and so the values of the parameters
$y_1$, $y_{12}$, $y_{123}$ do not matter. 

\section{Numerical study of two coupled models}
\label{sec5}

It was mentioned in the Introduction that the Potts model is usually
critical on the selfdual manifolds, for suitable values of $q$ (i.e.,
$0 \le q \le 4$). We expect this also to be true for coupled Potts models,
and so it is interesting to determine the corresponding universality
classes. In the lack of an exact (Bethe Ansatz) solution, this question
can be addressed by evaluating the effective central charge along the
selfdual manifolds, e.g., using numerical transfer matrix techniques.

In this Section we focus on the selfdual curve (\ref{N2sola}) for two
coupled models with pure two-spin interactions. Note that in
Section~\ref{sec2d} we have already remarked on a few special values
of the parameter $g$ for which the physics of the two coupled models
can be related to that of a single model.

\subsection{Transfer matrix}

The triangular-lattice Potts model can be transformed into a 
loop model on the medial (surrounding) graph---which is the Kagom\'e
lattice---in a standard way \cite{BaxterBook,Dotsenko99}. (This loop model
should not be confused with the one described in Section~\ref{sec3}.)
We have computed the free energy of the two coupled models (\ref{H2}) on
semi-infinite strips by constructing the transfer matrix of two
coupled Kagom\'e-lattice loop models.

\begin{figure}
\begin{center}
 \leavevmode
 \epsfysize=45mm{\epsffile{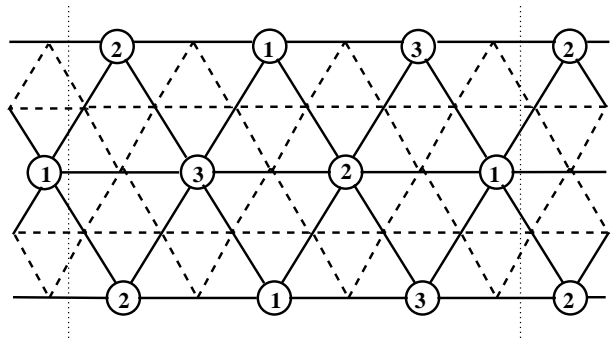}}
 \end{center}
 \protect\caption[3]{Semi-infinite strip, here of size $L=2$ triangles in the
 finite (vertical) direction. Periodic boundary conditions identify the top
 and the bottom of the figure. Potts spins are defined at the loci of the
 small circles; they interact along the solid lines which form a triangular
 lattice. The labels within each circle identify the usual three sublattices
 of the triangular lattice. The loop model is defined on the medial Kagom\'e
 lattice, shown in broken linestyle. The transfer matrix propagates the system
 along the horizontal direction, from left to right. Thin dotted lines
 indicate successive time slices (see text).}
\label{fig:kagome}
\end{figure}

The geometry is depicted in Fig.~\ref{fig:kagome}. For a periodic strip of
circumference $L$ triangles, each time slice cuts $2L$ dangling ends of the
Kagom\'e-lattice loop model. In order to have the leading eigenvalue
$\Lambda_0$ of the transfer matrix ${\bf T}_L$ correspond to the ground state
of the continuum model, the definition of ${\bf T}_L$ must respect the usual
sublattice structure of the triangular lattice. This means that $L$ must be
even, and that successive time slices are as shown in Fig.~\ref{fig:kagome}.

The numerical diagonalisation is most efficiently performed by decomposing
${\bf T}_L$ in a product of sparse matrices, each adding one vertex of the
Kagom\'e lattice (or, equivalently, one edge of the triangular lattice). With
the setup of Fig.~\ref{fig:kagome}, all these sparse matrices are identical,
except for the position of the two dangling ends on which they act. We have
been able to diagonalise ${\bf T}_L$ for sizes up to $L=10$ (the corresponding
matrix has dimension $141\,061\,206$).

\subsection{Central charge}

The free energy per unit area is $f(L)=-\frac{1}{4 \sqrt{3} L}\log \Lambda_0$,
the length scale being the height of one triangle. We have extracted
values of the effective central charge $c$ from three-point fits of the
form \cite{central}
\beq
 f(L) = f(\infty) - \frac{\pi c}{6 L^2} + \frac{A}{L^4},
 \label{3pfits}
\eeq
where the non-universal term in $A$ is supposed to adequately represent
the higher-order corrections.

\begin{figure}
\begin{center}
 \leavevmode
 \epsfysize=120mm{\epsffile{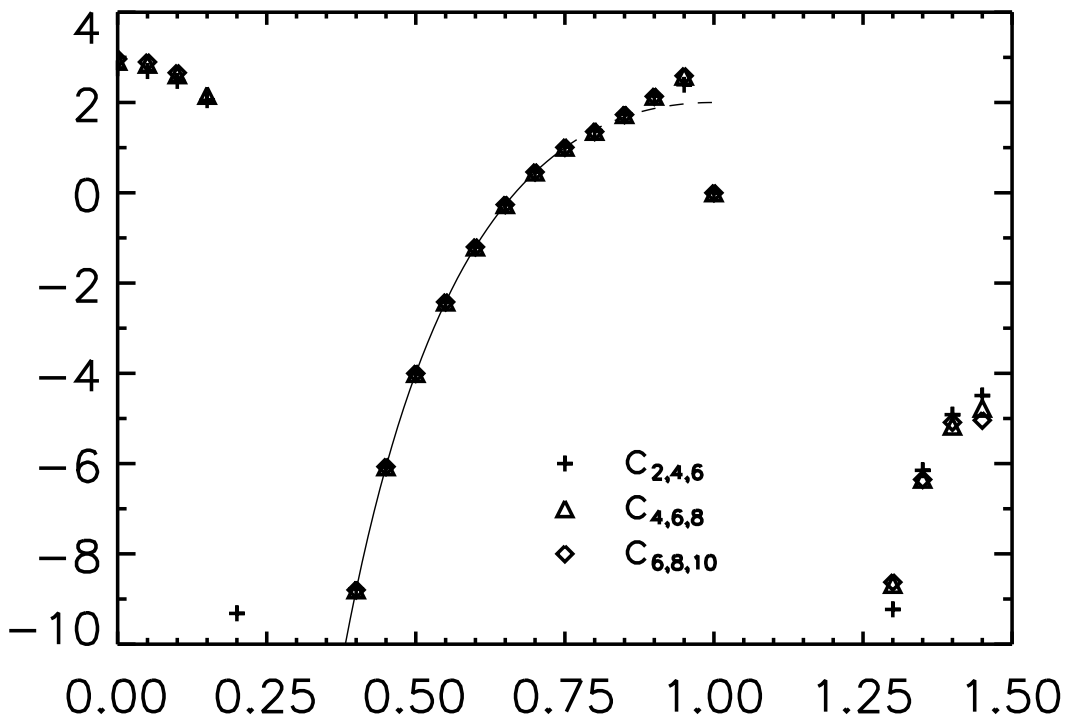}}
 \end{center}
 \protect\caption[3]{Three-point fits for the effective central charge $c$
 as a function of $g$. The solid curve shows the exact result
 (\ref{aprediction}), valid for $\frac14 \le g < \frac34$, as discussed in
 the text.}
\label{fig:central}
\end{figure}

Fig.~\ref{fig:central} shows the numerical values of $c(g)$ along the curve
(\ref{N2sola}). For each value of $g$, three estimates for $c(g)$
are shown, obtained by fitting $\{ f(L-4),f(L-2),f(L) \}$ to (\ref{3pfits})
for $L=6$, $L=8$ and $L=10$ respectively.

Naively, one would expect the $K_{12}$ coupling to be marginal at the
$q=2$ ferromagnetic point (i.e., at $g=\frac34$) and the surrounding
regime to be accessible to perturbative calculations. However, it
should be remembered that 1) the point $g=\frac34$ is not that of two
decoupled Ising models but that of a single 4-state model, and that 2)
the renormalisation group equations for $N$ coupled models are
singular when $N=2$ \cite{Dotsenko99}. Nevertheless, the numerics seems
quite conclusive that the $q < 2$ regime with $\frac14 \le g < \frac34$
has a central charge which is just twice that of (\ref{c1model}) [upon
changing the parametrisation, $g \to 1-g$]:
\beq
 c(g) = 2 \left(1 - \frac{6(1-g)^2}{g} \right), \qquad
 \mbox{for } \frac14 \le g < \frac34,
 \label{aprediction}
\eeq
meaning presumably that the continuum limit is really that of two decoupled
models. This is also consistent with the result $c(g=\frac12)=-4$. The
agreement of the numerics with (\ref{aprediction}) is excellent also for those
data points (with $0.25 \le g \le 0.35$) which are not visible in
Fig.~\ref{fig:central}.

The region $0 \le g \le g_1$, with $g_1 \approx 0.15$, is interesting
as it corresponds to $c>2$. This hints at the coupled models requiring
some kind of higher symmetry than its two constituent bosonic
theories.  Note in particular that for $g=0$, our three estimates for
$c$ read $c_{2,4,6}=2.758$, $c_{4,6,8}=2.914$ and $c_{6,8,10}=2.966$,
which we extrapolate to $c(g=0) = 3.00 \pm 0.01$. We conjecture that
the exact result is $c(g=0) = 3$. Since $q=4$, this
theory can also be represented as two coupled vertex models on the
Kagom\'e lattice \cite{BaxterBook}.

For $g=1$, the Boltzmann weights (\ref{Boltzmann}) are all $\pm 1$. It turns
out that in this particular case it is more convenient to work with a modified
transfer matrix that adds not one but $L/2$ time slices,
cf.~Fig.~\ref{fig:kagome}. This matrix has its largest eigenvalue equal to
unity regardless of $L$, and we conclude that $f(L)=0$ for any $L$. In
particular, this means $c(g=1) = 0$. The numerics is however indicative of a
non-trivial regime for $\frac34 < g < 1$, and it seems that we may have
$c(g)\to 4$ as $g \to 1^-$, consistent with two decoupled models each of
which is obtained by taking the limit $g \to 1^+$ in (\ref{c1model}).

In the region $1 < g < g_2$, with $g_2 \approx 1.10$, our numerical
diagonalisation scheme experiences difficulties, maybe due to the leading
eigenvalue having a non-zero imaginary part.

Finally, in the regime $g_2 < g < \frac32$ the central charge takes
large, negative values (in particular, some of the values are not
visible in Fig.~\ref{fig:central}). At first sight one might believe
that the continuum limit is that of two decoupled models in the
Berker-Kadanoff phase, i.e., with $c(g)$ given by twice that of
(\ref{c1model}) [upon changing the parametrisation $g \to 2-g$]:
$c(g) = 2 \big( 1 - 6 (2-g)^2/(g-1) \big)$. This possibility is
however clearly ruled out by the numerics, and $c(g)$ appears to be
given by a non-trivial expression.

As $g \to \frac32$, the two leading eigenvalues of the transfer matrix become
degenerate. In the sense of analytically continuing the curve (\ref{N2sol})
to negative values of $q$, this presumably marks a transition to non-critical
behaviour for $q<0$, i.e., with the phase transitions being first-order
in $b$ upon crossing the curve (\ref{N2sol}).

\section{Conclusion}
\label{sec6}

Using a mapping of coupled Potts models on the triangular lattice to
coupled loop models, we have obtained non-trivial selfdual manifolds
for two and three coupled Potts models with two and three-spin interactions.
A numerical study of the case of two coupled models shows that these
manifolds are good candidates for novel critical points, in particular in the
antiferromagnetic and unphysical regimes.

The technique can be applied to any number of coupled models, but expressing
the solutions explicitly in terms of the original coupling constants becomes
increasingly complicated as the number of models grows. This is in contrast
to the quite simple results for coupled Potts models on the square lattice
\cite{Jacobsen00,Dotsenko02}.

It would be interesting to study the simplest non-trivial case (\ref{N2sola})
using the methods of integrable systems.

\section*{Acknowledgments}

We would like to thank Henk J.~Hilhorst for an interesting comment which
motivated us for undertaking the present study.

\end{document}